# EPOBF: ENERGY EFFICIENT ALLOCATION OF VIRTUAL MACHINES IN HIGH PERFORMANCE COMPUTING CLOUD

**Nguyen Quang-Hung, Nam Thoai, Nguyen Thanh Son**

*Faculty of Computer Science & Engineering, HCMC University of Technology, VNUHCM, Vietnam*

[*]Emails: *{hungnq2,nam}@cse.hcmut.edu.vn*

## ABSTRACT

Cloud computing has become more popular in provision of computing resources under virtual machine (VM) abstraction for high performance computing (HPC) users to run their applications. A HPC cloud is such cloud computing environment. One of challenges of energy-efficient resource allocation for VMs in HPC cloud is trade-off between minimizing total energy consumption of physical machines (PMs) and satisfying Quality of Service (e.g. performance). On one hand, cloud providers want to maximize their profit by reducing the power cost (e.g. using the smallest number of running PMs). On the other hand, cloud customers (users) want highest performance for their applications. In this paper, we focus on the scenario that scheduler does not know global information about user jobs and/or user applications in the future. Users will request short-term resources at fixed start-times and non-interrupted durations. We then propose a new allocation heuristic (named Energy-aware and Performance-per-watt oriented Best-fit (EPOBF)) that uses metric of performance-per-watt to choose which most energy-efficient PM for mapping each VM (e.g. maximum of MIPS/Watt). Using information from Feitelson's Parallel Workload Archive to model HPC jobs, we compare the proposed EPOBF to state-of-the-art heuristics on heterogeneous PMs (each PM has multicore CPU). Simulations show that the EPOBF can reduce significant total energy consumption in comparison with state-of-the-art allocation heuristics.

*Keywords:* energy-aware scheduling, VM allocation.

## 1. INTRODUCTION

Cloud data centers use virtualization technology for provision computational resources in the form of virtual machines (VMs). Saving operating costs in terms of energy consumption (Watt-Hour) for a cloud system is highly motivated for any cloud resource owner. Energy-efficient management for HPC cloud was concerned some years ago [1 – 3]. The Green500 list, which has been presented since 2006 [4], has become popular. The Green500 list's idea ranks HPC systems based on a metric of performance-per-watt (FLOPS/Watt), implying that the higher FLOPS/Watt, the more energy-efficient HPC system.

Energy-efficient management for HPC cloud is still challenging. One of the challenges of energy-efficient scheduling algorithms is trade-off between minimizing energy consumption and satisfying Quality of Service (e.g. performance or resource availability on time for any reservation request). Resource requirements are application-dependent. However, HPC applications are mostly CPU-intensive and as a result, there could be unsuitable for dynamic consolidation and migration techniques as shown in [5] on HPC jobs/applications to reduce energy consumption of physical machines (PMs). Inspired by the Green500 list's idea [4], in this paper, we propose new VM allocation heuristics (named EPOBF) that use similar metric of *performance-per-watt* to choose the most energy-efficient PM for mapping each VM. We propose two methods to calculate the performance-per-watt values. We have implemented the EPOBF heuristics as extra VM allocation heuristics in the CloudSim version 3.0 [6]. We compare the proposed EPOBF heuristics to a popular VM allocation heuristic which is PABFD (Power-Aware Best-fit Decreasing) [5] and vector bin packing greedy L1/L2 (VBP Greedy L1/L2) heuristics. The PABFD [5] is a best-fit heuristic to choose which PM has least increasing power on placement of each VM. The VBP Greedy L2/L1 is a Norm-based Greedy L2/L1 in [7]. We evaluate these heuristics by simulations with a large-scale simulated system model, which has 5000 heterogeneous PMs and simulated workload with 29624 cloudlets each of which can model a HPC task. These simulated cloudlets use information that is converted from a Feitelson's Parallel Workload Archive [8] (SDSC-BLUE-2000-4.1-cln.swf [9]) to model HPC workload. Simulations show that the EPOBF heuristics can reduce 35 % total energy consumption on average in comparison to the PABFD and VBP Greedy allocation heuristics.

## 2. RELATED WORK

B. Sotomayor et al. [10][11], which is one of the earliest works on resource provision under VM abstraction in virtualized data centers, presented a lease-based model which uses leases and VMs abstraction for user short-term needs. They developed First Come First Serve (FCFS) and greedy mapping algorithms to map leases that included some of VMs with/without start time and user specified durations to a set of homogeneous physical machines (PMs). On one hand, the greedy algorithm can allocate a lease on the best performance physical machine to maximize performance. On the other hand, the greedy algorithm can allocate a small lease (e.g. with one VM) to a multicore physical machine. As a result, the greedy algorithm cannot optimize for energy efficiency.

Energy-aware resource management for a HPC data center is critical. Takouna, I. et. al., [3] presented power-aware multicore scheduling and FindHostForVm to select which host has minimum increasing power consumption to assign a VM. The FindHostForVm, however, is similar to the PABFD's [5] except that they concern memory usage in a period of estimated runtime for estimating host energy. The work also presented a method to select optimal operating frequency for a (DVFS-enabled) host and configure the number of virtual cores for VMs. Our EPOBF's FindHostForVm is different from the previous works in a way that the EPOBF's FindHostForVm chooses which host has the highest value of ratio of total maximum of MIPS (in all cores) to the host's maximum value of power consumption.

Energy-efficient job scheduling algorithms have been in an active research area recently. Albers, S. et. al. [12] reviewed some energy efficient algorithms which were used to minimize flow time by changing processor speed adapted to job size. G. Laszewski et al. [13] proposed scheduling heuristics and presented application experiences for reducing power consumption of parallel tasks in a cluster with the Dynamic Voltage Frequency Scaling (DVFS) technique. In this paper, we do not use the DVFS technique to reduce energy consumption on a cloud data center.

Mämmelä, O. et. al., [2] presented energy-aware First-In, First-Out (E-FIFO) and energy-aware Backfilling First-Fit (E-BFF, E-BBF) scheduling algorithms for non-virtualized high performance computing system. The E-FIFO puts new job at the end of job-queue (and dequeue last), finds out an available host for the first job and turns off idle hosts. The E-BFF and E-BBF are similar to E-FIFO, but the E-BFF and E-BBF will attempt to assign jobs to all idle hosts. Unlike our proposed EPOBF, these scheduling algorithms do not consider power-aware VM allocation.

Power-aware VM allocation heuristics for energy-efficient management in cloud data center were studied [14][5][15][16]. A. Beloglazov et al. [14][5][15] proposed the VM allocation as a bin-packing problem. They have presented a best-fit decreasing heuristic on VM allocation, named PABFD (power-aware best-fit decreasing) [5], and VM migration policies under lower and upper thresholds [14]. The PABFD prefers to allocate a VM to a host that will increase least power consumption. The PABFD can assign VMs to a host that has a few cores and authors concerns on only CPU utilization. Í. Goiri et. al. [16] has developed a score-based scheduling which is hill-climbing algorithm search for best match <host, VM> pairs. In which, score of each <host,VM> pair is sum of many factors such as power consumption, hardware and software fulfillment, resource requirement. Differently, the both EPOBFs choose a host that has maximum of MIPS/Watts to assign a VM. We concern on three resource types are processing power (e.g. MIPS), size of physical memory and network bandwidth and energy consumption. These studies can be suitable for service allocation, in which each VM will execute a long running, persistent application. Instead, we consider provision of resources for HPC applications that will start at a fixed point in time for a non-interrupted duration. Finally, the aforementioned points make our paper distinguished from the previous works reviewed in [17][18][19].

## 3. PROBLEM FORMULATION

### 3.1. Terminology and notations

We describe notation used in this paper as following:

| Notation | Description |
|----------|-------------|
| $VM_i$ | the i-th virtual machine |
| $M_j$ | the j-th physical machine (host) |
| $r_j(t)$ | set of indexes of virtual machines that is allocated on the $M_j$ at time $t$ |
| $U_{cpu}(t)$ | CPU utilization of a physical machine at time $t$, $0 \leq U_{cpu}(t) \leq 1$ |
| $mips_{i,c}$ | allocated MIPS of the $c$-th processing element (PE) to the $VM_i$ by $M_j$ |
| $MIPS_{j,c}$ | total MIPS of the $c$-th processing element (PE) on the $M_j$ |

## 3.2. Power Model

We assume that total power consumption of a single physical machine ($P(.)$) has a linear relationship with CPU utilization as mentioned in [20]. We calculate CPU utilization of a host is sum of total CPU utilization on $PE_j$ cores:

$$U_{cpu}(t) = \sum_{c=1}^{PE_j} \sum_{i \in r_j(t)} \frac{mips_{i,c}}{MIPS_{j,c}} \qquad (1)$$

Total power consumption of a single host ($P(.)$) at time $t$ is calculated:

$$P\big(U_{cpu}(t)\big) = P_{idle} + (P_{max} - P_{idle}).U_{cpu}(t) \qquad (2)$$

Total energy consumption of a host ($E$) in period time [$t_1$, $t_2$] is defined by:

$$E = \int_{t1}^{t2} P\big(U_{cpu}(t)\big) dt \qquad (3)$$

We assume that a cloudlet [6] is executed only in a single VM and both VM and cloudlet terminate simultaneously. The CloudSim [6] only calculates energy consumption on any host that is executing at least one VM (i.e. total of CPU utilization of the host is greater than zero).

## 3.3. Problem Description

We consider the problem of energy-efficient allocation of VMs in HPC Cloud. We formulate the scheduling problem as following:

"Given a set of $n$ virtual machines {$VM_i(pe_i, mips_i, ram_i, bw_i, ts_i, d_i)$ |$i = 1,...,n$} to be placed on a set of $m$ heterogeneous physical machines {$M_j(PE_j, MIPS_j, RAM_i, BW_j)$ |$j = 1,...,m$}. Each virtual machine $VM_i$ requires $pe_i$ processing elements, $mips_i$ MIPS, $ram_i$ MBytes of physical memory, $bw_i$ Kbits/s of network bandwidth, and the $VM_i$ will be started at time ($ts_i$) and finished at time ($ts_i + d_i$) without neither preemption nor migration in its duration ($d_i$). We concern three types of computing resources such as processors, physical memory, and network bandwidth."

We assume that every host $M_j$ can run any VM and power consumption model ($P_j(t)$) of the $M_j$ has a linear relationship with CPU utilization as in formula (2).

The objective scheduling is trade-off between minimizing total energy consumption in fulfillment of maximum requirements of $n$ VMs and following constraints:

Constraint 1: A virtual machine (VM) runs only on identified host, except that the VM is being migrated. However, we do not consider on migration of VM in this paper.

Constraint 2: No VM requests any resource is larger than capacity of the resource in a host.

Constraint 3: Let $r_j(t)$ be set of indexes of VMs that are allocated to a host $M_j$. There is sum of total demand resource of these allocated VMs is less than or equal to maximum capacity of the resource of the $M_j$. For each $c$-th processing element of physical machine $M_j$ ($j = 1,..,m$):

$$\forall c = 1 \dots PE_j, \forall i \in r_j(t): \sum_{i \in r_j(t)} mips_{i,c} \le MIPS_{j,c}$$

For other resources of the $M_j$ such as physical memory (RAM) and network bandwidth (BW):

$$\forall i \in r_j(t): \sum_{i \in r_j(t)} ram_i \le RAM_j, \forall i \in r_j(t): \sum_{i \in r_j(t)} bw_i \le BW_j$$

HPC applications have various sizes and require multiple cores and submit to system at dynamic arrival rate [21]. A HPC application can request some VMs.

# 4. ENERGY-AWARE SCHEDULING ALGORITHM

Inspired by the Green500 list's idea [4] that uses a metric of performance-per-watt (FLOPS/watt) to rank energy efficiency of HPC systems, we raise questions: how can we use a similar metric (e.g. TotalMIPS/Watt) as a criterion for selecting a host on assignment of a new VM and is total energy consumption of the whole system minimum?

We assume that if a host has more number of cores then the host will have more number of MIPS/Watt. The number of MIPS/Watt of a host is a ratio of total maximum of MIPS, which is sum of total maximum of MIPS of all host's cores, to its maximum power consumption ($P_{max}$). The objective of our proposed work is energy efficiency. We present here the EPOBF with two ways to calculate the metric of performance-per-watt.

**EPOBFv1**: The EPOBF will sort list of VMs and order them by earliest start time first, then the EPOBF will assign a VM to a host has the highest $G$ value. For any host $h$, the $G_h$ can be calculated as a ratio of total maximum of MIPS of the host $h$ (sum of total MIPS of all cores) to maximum power consumption at 100% CPU utilization ($P_{max}$) of the host $h$. We called the $G$ is in metric of performance-per-watt. In summary, the EPOBF assigns each ready virtual machine $v$ to the host $h$ has maximum of the $G_h$ values. Algorithm 1 below shows pseudo-code for the EPOBF.

---

**Algorithm 1: EPOBFv1** (Energy-aware Performance-per-watt Oriented Best-Fit)

**Input:** list of VMs (V), list of hosts (H)
**Output**: a mapping of the list of VMs
    Sort the list of VMs order by their earliest starting time first.
    ForEach (vm in V):
        host = FindHostForVmByGreenMetric(vm, H)
        CloudSim.allocationMap.put(vm.getId(), host.getId());
    End For
    Return true

**FindHostForVmByGreenMetric:**
**Input:** a VM (vm) and a set of hosts (H)
**Output**: bestHost - a best host for allocation of the *vm*
    Hosts = FindCandidateHosts(vm, H)
    Forall (PowerHost $h$ in Hosts ):

$$G_h = \frac{h.TotalMIPS}{h.GetPower(100\%)} \quad\quad (4)$$

    bestHost $\leftarrow$ { Choose which host $h$ has the $G_h$ value is maximum }.
    Return bestHost

---

**EPOBFv2**: The EPOBFv2 is similar to the EPOBFv1, except that the EPOBFv2 modifies the way to calculate the $G_h$ value for a host $h$. For a host $h$, the $G_h$ can be calculated as a ratio of total maximum of MIPS (sum of total MIPS of all cores) of the host $h$ to increasing power consumption after allocation of the virtual machine ($vm$) that is assigned to the host $h$. For each host $h$, the $G_h$ value of the host $h$ is calculated as follows:

$$G_h = \frac{h.TotalMIPS}{h.CurrentPower - h.GetPowerAfterAllocateVm(vm)}. \quad\quad (5)$$

## 5.  SIMULATION AND EVALUATION

In this section, we discuss about statistical analysis for the simulated workload that is used in the simulations using CloudSim [6]. Then we discuss the results of the proposed EPOBFv1 and EPOBFv1 in comparison with other heuristics in the existing works.

### 5.1.  Statistical Analysis of Simulated Workload

We evaluate these heuristics by simulations with a simulated cloud data center system that has 5000 heterogeneous hosts (there are three groups of hosts with different cores and physical memory), and a simulated workload with 29624 CloudSim's cloudlet  [6] (we assume that each HPC job's task is modeled as a cloudlet). The workload's information of tasks is extracted from a real log-trace (SDSC-BLUE-2000-4.1-cln.swf [9]) in Feitelson's Parallel Workloads Archive (PWA) [8] to model HPC jobs (each job must have executing time at least 300 seconds). When converting from the log-trace, each cloudlet's length is a product of the system's processing time and CPU rating (we set the CPU rating is 375). We assign job's submission time, start time, and execution time, and number of process in job information in the SDSC-BLUE-2000-4.1-cln to cloudlet's submission time, start time and cloudlet's length, and number of cloudlets. Figure 1 shows histogram of start-time rate and length in millions of instructions of the 29624 cloudlets.

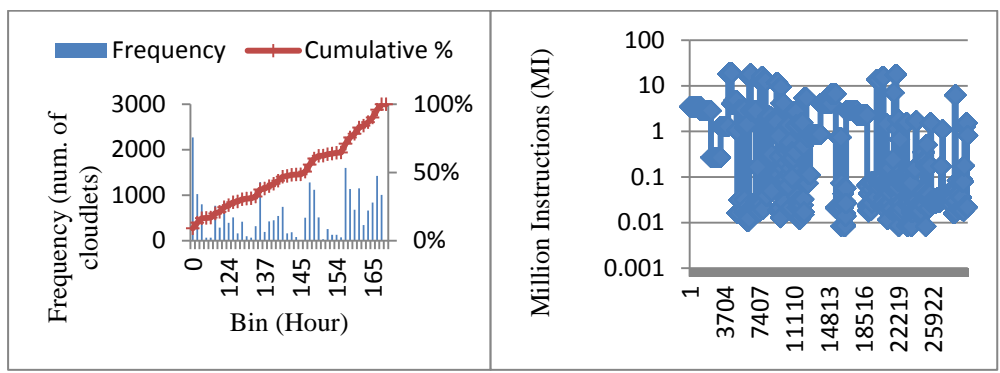

*Figure 1.* Histogram of start-time rate of the 29624 simulated cloudlets (left chart) and length in Millions Instruction (MI) of the 29624 simulated cloudlet (right chart).

### 5.2.  Simulations

#### 5.2.1. Methodology

We choose a latest version (version 3.0) of the CloudSim [6] to model and simulate our HPC cloud and the VM allocation heuristics. The CloudSim's framework is used to write user customized dynamic allocation algorithms for VMs and cloudlets. We based on the previous work [5] to improve existing PABFD allocation heuristic for this paper.

#### 5.2.2. Workload for simulations

We show here the results of simulations from simulated workload (with 29624 cloudlets) described above. We use four types of VMs (e.g. Amazon EC2's VM instances: high-CPU, extra, small, micro): each VM has only one core and maximum performance of VM is {2500, 2000, 1000, 500} MIPS, {870, 3840, 1740, 613} MB of RAM and network of 10000 Kbits/s.

*5.2.3. Simulated Cloud data center*

Our simulated Cloud data center has total 5000 heterogeneous PMs to evaluate our proposed VM allocation heuristics. These PMs include three groups of machines: one-third of HP ProLiant ML110 G5 machines, one-third of IBM x3250 machines, and one-third of Dell PowerEdge R620 machines. We assume that power consumption of a PM has a linear relationship to CPU utilization (Sec. 3.2). We use three power models of the three mainstream servers as summarized in Table 1:

*Table 1.* Server characteristics (A: HP ProLiant ML110 G5, B: IBM x3250, C: Dell PowerEdge R620).

| Server | A | B | C |
|---|---|---|---|
| CPU | 1x Xeon 3075 2.66GHz | 1x Xeon X3470 2.93GHz | 2x Xeon E5-2660 2.20GHz |
| Number of cores | 2 | 4 | 16 |
| RAM (GB) | 4 | 8 | 24 |
| Network BW (Kbits/s) | 10,000,000 | 10,000,000 | 10,000,000 |
| Maximum of MIPS/core | 2660 | 2933 | 2660 |
| $P_{idle}$ (Watt) | 93.7 | 41.6 | 56.1 |
| $P_{max}$ (Watt) | 135.0 | 113.0 | 263.0 |
| TotalMIPS/$P_{max}$ (MIPS/Watt) | 39.407 | 103.823 | 161.825 |

### 5.2.4. Results and Discussions

We compare the following VM allocation heuristics with ours, which have been defined in the previous section, on total energy consumption (kWh):

**PABFD (Power-Aware Best-Fit Decreasing):** The PABFD is presented in [5]. We use the PABFD as a baseline algorithm. The PABFD sorts the list of VMs in start time (i.e. earliest start time first) and uses a Best-Fit Decreasing heuristic by minimizing increasing power on each placement of VM.

**VBP greedy L2/L1**: We implemented two VBP Greedy L2/L1 packing heuristics that are presented as Norm-based Greedy L2/L1 in [7].

*Table 2.* Energy Consumption (kWh).

| VM Allocation Heuristic | #Hosts | #VMs | #Cloudlets | Energy (kWh) | Energy saving (+: better, -: worse) | #VM Migrations |
|---|---|---|---|---|---|---|
| PABFD | 5000 | 29624 | 29624 | 2513.38 | 0% | 0 |
| VBP Greedy L1 | 5000 | 29624 | 29624 | 8972.92 | -257% | 0 |
| VBP Greedy L2 | 5000 | 29624 | 29624 | 2513.38 | 0% | 0 |
| EPOBFv1 | 5000 | 29624 | 29624 | 1640.78 | 35% | 0 |
| EPOBFv2 | 5000 | 29624 | 29624 | 1640.78 | 35% | 0 |

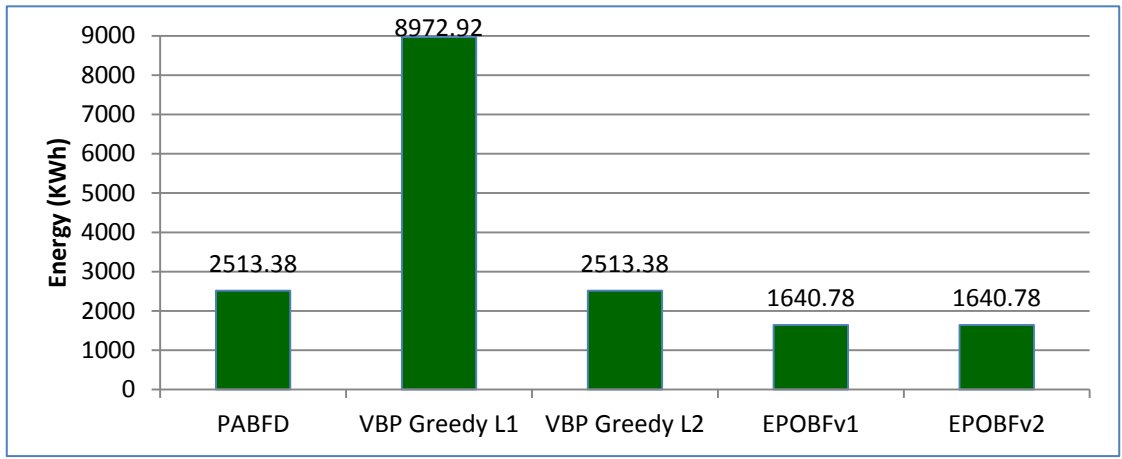

*Figure 2.* Energy consumption (KWh).

The Table 2 shows scheduling results from simulations. Figure 2 shows total energy consumption (kWh) of VM allocation heuristics: PABFD, VBP Greedy L[1-2] and EPOBFv[1-2], and Figure 3 shows percentages of energy savings of VBP Greedy L[1-2] and EPOBFv[1-2] in comparison with the PABFD, if the energy savings of a heuristic is a positive number, then the heuristic is better than the PABFD. Otherwise, the heuristic is worse than the PABFD.

Data in simulations (table 2) shows that compared with the PABFD and VBP Greedy L2 (VBP Greedy L1), the proposed EPOBFv1 and EPOBFv2 heuristics can reduce energy consumption to 35 % and 35 % (82 %) respectively. The EPOBF can reduce up to 35 % of total energy consumption because the EPOBF prefers to allocate a new VM to the Dell PowerEdge R620 machine. This mean the EPOBF chooses a PM with the maximum value of number of MIPS/Watt (*G* value). The PABFD chooses which host has the minimum value of increasing power consumption. Therefore the PABFD prefers to allocate a new VM to an IBM server x3250 machine, for example, given a set of 11 VMs (each VM needs 1 core) the PABFD will use three IBM server x3400 servers with 4 cores for mapping of 11 VMs, instead of that the EPOBF uses only one Dell PowerEdge R620 server with 16 cores. Such a choice makes our work more energy-efficient.

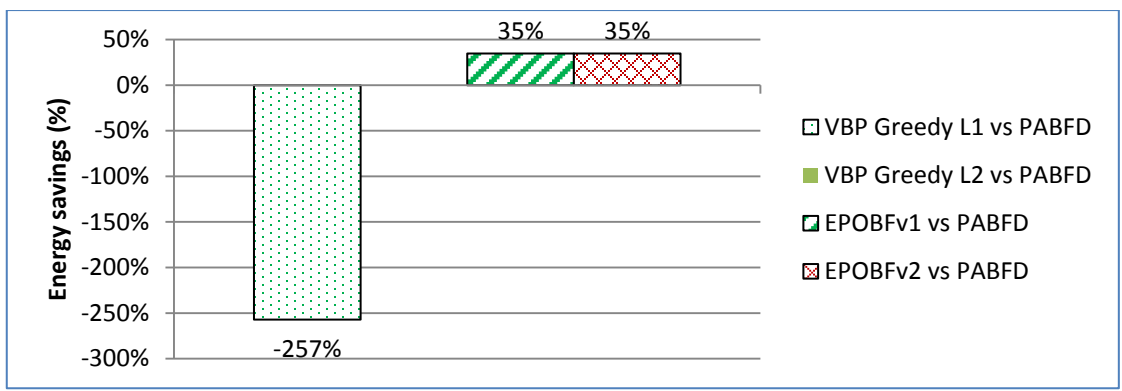

*Figure 3.* Energy savings (%) of VBP Greedy L[1-2] and EPOBFv[1-2] in comparison to PABFD.

## 6. CONCLUSIONS AND FUTURE WORKS

This work presented energy-efficient scheduling of VMs by EPOBFv1 and EPOBFv2 heuristics that can be applied to HPC clouds. A HPC cloud's scheduler can use the metric of performance-per-watt to allocate VMs to hosts for more energy efficiency. The experimental simulations show that we can reduce 35 % total energy consumption in comparison with the state-of-the-art VM allocation heuristics (e.g. PABFD). The EPOBF heuristic could be a new VM allocation solution in a Cloud data center with heterogeneous and multicore physical machines. The both EPOBFs are better than the PABFD and VBP greedy L1/L2 allocation heuristics.

Evaluating the performance of the EPOBF heuristics on various system models and workloads to provide a pros and cons of the EPOBF heuristics is needed as our future work. Furthermore, we will concern deadlines for tasks running on the VMs; consider the impact of memory in energy models. An accurate power model for multicore PMs will also be investigated.

*Acknowledgments*. We acknowledge the support from the project "Developing a high performance computing cloud for small and medium scientific research groups" funded by VNU-HCM.